\documentclass[pra,superscriptaddress,twocolumn,showpacs]{revtex4}
\usepackage{amsmath,graphicx,epsfig,color,amsfonts}
\usepackage[hypertex]{hyperref}
\usepackage{amsthm}

\newcommand{\ket}[1]{|#1\rangle}

\newcommand{\unit}{\mathbf{1}}
\renewcommand{\S}{\mathcal{S}}

\newcommand{\St}{\S^{(\tau)}}
\newcommand{\StObs}{\S^{(\tau)}_{\mbox{\tiny Obs}}}
\newcommand{\SCHObs}{\S^{(\mbox{\tiny CH})}_{\mbox{\tiny Obs}}}
\newcommand{\SCH}{\S^{(\mbox{\tiny CH})}}
\newcommand{\It}{I^{(\tau)}}
\newcommand{\Sq}{\S_{\mbox{\tiny qm}}}
\newcommand{\PS}[2]{\ket{\psi^{#1}_{#2}}}

\newcommand{\GPSqb}{\PS{\gamma}{2}}
\newcommand{\PsiCr}{\ket{\psi^\tau_c}}
\newcommand{\Sqt}{\Sq^{(\tau)}}
\newcommand{\SqCH}{\Sq^{(\mbox{\tiny CH})}}
\newcommand{\Slhv}{\St_\text{sep}}
\newcommand{\A}{\mathcal{A}}
\newcommand{\B}{\mathcal{B}}
\newcommand{\sA}{x}
\newcommand{\sB}{y}
\newcommand{\oA}{a}
\newcommand{\oB}{b}
\newcommand{\half}{\frac{1}{2}}
\newcommand{\Coeff}{\beta_{\sA\sB}^{\oA\oB}}
\newcommand{\JProb}[4]{p(#2,#4|#1,#3)}

\newcommand{\MProbA}[2]{p_\A(#2|#1)}
\newcommand{\MProbB}[2]{p_\B(#2|#1)}
\newcommand{\GJProb}{\JProb{\sA}{\oA}{\sB}{\oB}}
\newcommand{\GMProbA}{\MProbA{\sA}{\oA}}
\newcommand{\GMProbB}{\MProbB{\sB}{\oB}}
\newcommand{\GMProbAObs}{p_{\mbox{\tiny Obs}}(\oA | \sA)}
\newcommand{\GMProbBObs}{p_{\mbox{\tiny Obs}}(\oB | \sB)}

\newcommand{\AllProb}{\{\GJProb\}_{x,y,a,b\in\{0,1\}}}
\renewcommand{\t}{^{\mbox{\tiny T}}}

\newcommand{\GPOVMA}{A_{\sA}^{\oA}}
\newcommand{\GPOVMB}{B_{\sB}^{\oB}}

\newcommand{\proj}[1]{\left| #1 \right\rangle\!\!\left\langle #1 \right|}
\DeclareMathOperator{\tr}{tr}

\newtheorem{theorem}{Theorem}
\newtheorem{conjecture}{Conjecture}

\begin{document}

\title{Semi-device-independent bounds on entanglement }

\author{Yeong-Cherng~Liang}
\email{yeongcherng.liang@unige.ch}
\affiliation{Group of Applied Physics, University of Geneva, CH-1211 Geneva 4, Switzerland}
\affiliation{School of Physics, University of Sydney, New South Wales 2006, Australia.}

\author{Tam\'as V\'ertesi}
\affiliation{Institute of Nuclear Research of the Hungarian
Academy of Sciences, H-4001 Debrecen, P.O. Box 51, Hungary.}

\author{Nicolas Brunner}
\affiliation{H.H. Wills Physics Laboratory, University of
Bristol, Bristol, BS8 1TL, United Kingdom.}

\date{\today}

\begin{abstract}
Detection and quantification of entanglement in quantum resources are two key steps in the implementation of
various quantum-information processing tasks. Here, we show that Bell-type inequalities are not only useful in
verifying the presence of entanglement but can also be used to bound the entanglement of the underlying physical
system. Our main tool consists of a family of Clauser-Horne-like Bell inequalities that cannot be violated
maximally by any finite-dimensional maximally entangled state. Using these inequalities, we demonstrate the
explicit construction of both lower and upper bounds on the concurrence for two-qubit states. The fact that these
bounds arise from Bell-type inequalities also allows them to be obtained in a semi-device-independent manner,
that is, with assumption of the dimension of the Hilbert space but without resorting to any knowledge of the
actual measurements being performed on the individual subsystems.
\end{abstract}

\maketitle

\section{Introduction}\label{Sec:Introduction}

Entanglement~\cite{Horodecki:RMP:entanglement} has long played a pivotal role in many quantum information processing tasks, such as secure communication using quantum-key distribution~\cite{A.K.Ekert:PRL:1991,BHK,NLcrypto,DIQKD}, teleportation of quantum states~\cite{Teleportation}, quantum computation~\cite{Noah.Linden:PRL:2001}, reduction in communication complexity~\cite{CC}, and more recently, expansion and certification of randomness~\cite{BIV:Randomness,R.Colbeck:thesis}. As a result, the verification and quantification of this resource present in quantum systems is often an indispensable part of quantum-information processing (QIP) protocols.

Traditionally, for low-dimensional composite systems that are made up of only a few subsystems, the process of entanglement certification and/or quantification is carried out using complete quantum state tomography followed by the application of certain separability criteria~\cite{O.Guhne:PR:2009}. This approach, however, suffers from the drawbacks that it is unnecessarily resource intensive and that the procedure of tomography may already be intractable for a physical system that is made up of only a handful of qubits (see, for example, Ref.~\cite{TomographyHard} and references therein).

More recently, there has been considerable interest in verifying the presence of entanglement by measuring directly the expectation value of so-called  {\em entanglement witnesses}~\cite{Horodecki:RMP:entanglement,O.Guhne:PR:2009}. While this latter approach is much more economical in terms of both the number of measurement settings and the resource required for classical post-processing \cite{EW}, it still requires detailed knowledge of the actual measurements performed on the physical system in question. This last feature is, of course, undesirable  as there are examples of entanglement witnesses that are extremely fragile to the perturbations of the constituent measurement operators~\cite{DIEW}.  Also, in some QIP tasks such as quantum key distribution, it is highly desirable to make minimal assumptions on the devices used in the protocol.

In contrast, violation of Bell-type inequalities~\cite{Bell} is only a statement about the measurement statistics derived from an experiment. Since entanglement is a prerequisite for Bell-inequality violation~\cite{R.F.Werner:PRA:1989}, one can deduce the presence of entanglement via a Bell-inequality violation without resorting to any knowledge of the actual physical implementation of the measurements. In modern terminology, Bell-type inequalities therefore serve as {\em device-independent entanglement witnesses} and can be used for device-independent quantum-key distribution \cite{DIQKD,S.Pironio:NJP:diqkd}, state estimation \cite{DISE}, randomness generation \cite{BIV:Randomness,R.Colbeck:thesis}, as well as self-testing quantum circuits \cite{mayers,mckague}.

For a long time, it was thought that the strength of Bell inequality violation is monotonously related to the degree of entanglement (see, e.g., Refs.~\cite{R.F.Werner:QIC:2001,Y.C.Liang:Thesis} for a review on this topic); that is, loosely speaking, more entanglement would lead to more nonlocality. It turns out that the relation between entanglement and nonlocality is much more intricate \cite{methotscarani}. However, for the Clauser-Horne-Shimony-Holt (CHSH) Bell inequality \cite{Bell:CHSH}, it was shown by Verstraete and Wolf \cite{F.Verstraete:PRL:170401} that the {\em maximal} possible CHSH violation of a two-qubit state is monotonously related to the entanglement of the latter. From this result, it is then easy to lower bound the entanglement of a two-qubit state given the observed CHSH violation (see also Ref.~\cite{DISE}).

On the other hand, Bell inequality violation --- as we demonstrate in this paper --- can also be used to upper bound the entanglement of the underlying state. To achieve this, we present and make use of a family of Clauser-Horne-like ~\cite{CH:PRD:1974} Bell inequalities (i.e., involving two binary measurements per party) that are violated maximally by non-maximally entangled states. Moreover, it can be shown that some of these inequalities {\em cannot} be violated maximally by maximally entangled states of any dimensions.
This complements the recent result of Junge and Palazuelos \cite{Junge:1007.3043}, who proved the existence of such inequalities.

Naturally, our inequalities suggest another {\em device-independent} application of Bell inequalities, namely, to bound---from above and below---the entanglement present in some underlying quantum system directly from the measurement statistics. However, in contrast with the scenario considered in {\em device-independent quantum-key distributions}~\cite{DIQKD,S.Pironio:NJP:diqkd}, here we assume that the dimension
of the underlying Hilbert space is known. To distinguish the present approach (where the dimension
of the Hilbert space is assumed) from the conventional device-independent approach, hereafter, we will refer to
our scenario as {\em semi-device-independent}. We focus on a scenario
where the experimentalists know on which degree of freedom of
the system the measurements are performed --- thus the relevant Hilbert
space dimension can be determined --- but nothing is assumed about the
alignment of the measuring devices. We believe this represents a
significant advantage compared to previous techniques, such as
entanglement witnesses and tomography, for which small alignment errors
can lead to important errors in the estimation of entanglement~\cite{DIEW}. We note also that it seems unlikely that one can find any useful bounds without such assumption on the dimensionality of the systems, as we discuss later.

The paper is organized as follows. In Sec.~\ref{Sec:Prelim}, we introduce a class of Bell inequalities that will  serve as our main tools for bounding entanglement. Then, in Sec.~\ref{Sec:MaxViolation}, we  investigate the quantum violations of these inequalities by maximally entangled states (the relevant proof is relegated to Appendix~\ref{App:Proofs}). After that, in Sec.~\ref{Sec:Bounds:Qubits}, we  illustrate how these Bell inequalities can be used to provide nontrivial lower and upper bounds on the entanglement of a two-qubit state. In Sec.~\ref{Sec:Qudits}, we  briefly comment on the applicability of these tools to the scenario where one makes no assumption of the dimension of the underlying quantum systems. We  conclude in Sec.~\ref{Sec:Conclusion} by comparing our approach with the one presented of Ref.~\cite{DISE} in the context of device-independent state estimation, and highlighting some possible avenues for future research. In Appendix~\ref{App:Projective}, we also provide another method of upper bounding entanglement assuming that the measurements are projective.

\maketitle

\section{Preliminaries}\label{Sec:Prelim}
In this section, we first introduce the notations and briefly review the tools that we are going to use in the subsequent analysis. To this end, we consider the simplest scenario where two experimenters Alice and Bob have access to many copies of a shared quantum state $\rho$ and after many rounds of experiment, the bipartite conditional probability distributions $\left\{\GJProb \right\}_{a,b,x,y\in\{0,1\}}$ are estimated from the experimental data. Here, $\GJProb$ refers to the joint probability that the outcomes $\oA$ and $\oB$ are observed at Alice's and Bob's site respectively, conditioned on her performing the $\sA^{\rm th}$ and him performing the $\sB^{\rm th}$ measurement. Clearly, from $\AllProb$, the respective marginal probabilities $\GMProbA$ and $\GMProbB$ can also be determined.  Our goal here is to obtain some non-trivial bounds on the entanglement of the underlying state from these measurement statistics.

Our starting point is the following one-parameter family of
modified Clauser-Horne (CH) polynomials:
\begin{subequations}\label{Eq:Bellfunction:tiltedCH}
\begin{equation}
    \St=\sum_{\sA,\sB,\oA,\oB=0, 1}
    \Coeff\, \GJProb,
\end{equation}
where
\begin{gather}\label{Eq:beta:explicit}
\beta_{0,1}^{0,0}=\beta_{1,0}^{0,0}=1-\tau,\quad
\beta_{0,1}^{0, 1}=\beta_{1,0}^{1,0}=-\tau,\\
\beta_{0,0}^{0,0}=-\beta_{1,1}^{0,0}=1,\quad
\beta_{\sA,\sB}^{a,b}=0\quad\text{otherwise},
\end{gather}
\end{subequations}
for $\tau\ge1$.
It is a well-known fact~\cite{R.F.Werner:PRA:1989} that with local measurements, separable states can only give rise to probability distributions that can be reproduced by {\em shared randomness}, and hence any measurement statistics that originate from such states must satisfy the Bell inequality
\begin{equation}\label{Ineq:tilted:CH}
    \It: \St=\Slhv\le 0.
\end{equation}
Note that this particular Bell inequality arises naturally in the study of the detection loophole in Bell experiments. Here the parameter $\eta=1/\tau$ can be interpreted as the (symmetric) detection efficiency~\cite{P.H.Eberhard:PRA:747,T.Vertesi:PRL:060401,Branciard:1010.1178}.

For $\tau=1$, the Bell inequality $\It$ reduces
to the CH inequality~\cite{CH:PRD:1974} which, in turn, is equivalent to the well-known CHSH inequality~\cite{Bell:CHSH}.
Thus, we can also rewrite
Eq.~\eqref{Eq:Bellfunction:tiltedCH} as
\begin{equation}
    \St=\S^{(\mbox{\tiny CH})}+
    (1-\tau)\left[\MProbA{0}{0}+\MProbB{0}{0}\right],
    \label{Eq:Bellfunction:tiltedCH:ViaCH}
\end{equation}
where $\SCH\equiv\S^{(\tau=1)}$ is the CH polynomial. From Eq.~\eqref{Eq:Bellfunction:tiltedCH:ViaCH}, it is clear that for any $\tau\ge 1$ we can see $\It$ as an inequality that consists of a conic combination of the CH inequality and some positivity constraints on joint probabilities. As a result, any measurement statistics that violate inequality~\eqref{Ineq:tilted:CH} with $\tau>1$ must also violate the CH inequality. 

Moreover, it is easily checked that any  legitimate probability distribution always satisfies inequality \eqref{Ineq:tilted:CH} for $\tau\geq \tfrac{3}{2}$,  since $\It$ corresponds to a conic combination of positivity constraints in these cases. 
Henceforth, we  restrict our attention to the scenarios where $1\le\tau<\tfrac{3}{2}$.

In the following discussion, we  denote by $\GPOVMA$ the positive-operator-valued
measure (POVM) element associated with the $\oA$-th outcome of Alice's
$\sA$-th measurement (likewise $\GPOVMB$ for Bob's). For
any quantum state $\rho$, we then write the corresponding quantum expectation
value of $\St$ under these measurement operators as:
\begin{equation}\label{Eq:Sqm:General}
    \Sqt\left(\rho,\left\{\GPOVMA, \GPOVMB\right\}\right)
    =\!\!\!\sum_{\sA,\sB,\oA,\oB} \Coeff
    \tr\left(\rho\, \GPOVMA\otimes \GPOVMB\right).
\end{equation}
The corresponding maximal quantum violation of
inequality~\eqref{Ineq:tilted:CH} for $\rho$ is denoted by:
\begin{equation}\label{Eq:MaxSqTau:FixedRho}
    \Sqt\left(\rho\right)=\max_{\GPOVMA,\GPOVMB}
    \Sqt\left(\rho,\left\{\GPOVMA, \GPOVMB\right\}\right),
\end{equation}
whereas the maximal violation of inequality~\eqref{Ineq:tilted:CH}
allowed by quantum mechanics is denoted by
\begin{equation}
    \Sqt=\max_{\rho} \Sqt\left(\rho\right).
\end{equation}

\section{Maximal violation by non-maximally entangled state}\label{Sec:MaxViolation}

\begin{figure}[b!]
    \includegraphics[width=9.0cm]{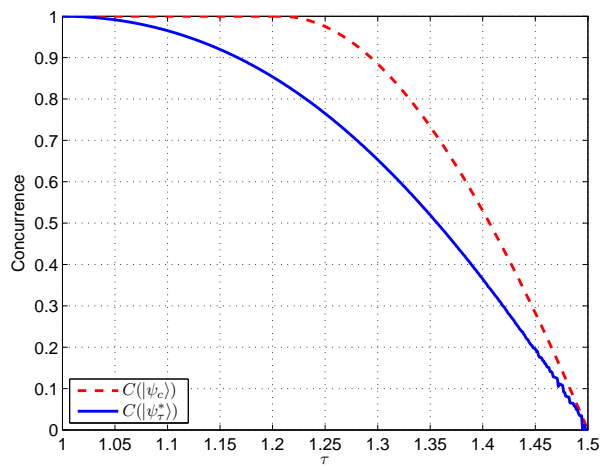}
    \caption{\label{Fig:ConcurrencevsTau}
(Color online) Plot of the concurrence $C$ of the two-qubit state $\ket{\psi^*_\tau}$ which violates maximally the inequality $\It$ as a
function of $\tau$ (blue solid line). Also plotted is the concurrence $C_\text{cr}$ of the critical two-qubit state $\PsiCr$ (red dashed line). All states with $C>C_{cr}$ do not violate $\It$ (see text at Sec.~\ref{Sec:UB:BIV} for details); hence, the violation of $\It$ allows one to derive upper bounds on the entanglement.}
\end{figure}

As mentioned in Sec.~\ref{Sec:Introduction}, an important feature of $\It$ that allows us to establish {\em upper} bound on the underlying entanglement is that, except for $\tau=1$, it is a Bell inequality that is violated maximally by non-maximally entangled two-qubit states (see Fig.~\ref{Fig:ConcurrencevsTau}). A natural question that one may ask is whether it is possible to recover the maximal quantum violation of $\It$ if we also consider maximally entangled states of higher dimension. We show in Appendix~\ref{App:MES:NMaxBIV} that this is not possible for a large range of $\tau$,  as summarized in the theorem below~\cite{fn:Filters}.

\begin{theorem}\label{Conjecture:MESCannotProduceQ}
No finite-dimensional maximally entangled state can violate maximally the Bell inequality
$\It$ with $\tau\ge\tfrac{1}{\sqrt{2}}+\tfrac{1}{2}$.
\end{theorem}

In relation to this, it is worth noting that for two-qubit states, one can show that the maximally entangled state cannot even violate $\It$ for the same interval of $\tau$ (let alone maximally). For higher dimensional maximally entangled states, despite intensive numerical search, we have also not found any violation of $\It$ with $\tau\ge\tfrac{1}{\sqrt{2}}+\tfrac{1}{2}$ for local Hilbert space dimension up to $d=50$. We believe that this feature holds for maximally entangled states of arbitrary dimensions~\cite{fn:DetectionLoophole}.

\begin{conjecture}
No finite-dimensional maximally entangled state can violate
$\It$ with $\tau\ge\tfrac{1}{\sqrt{2}}+\tfrac{1}{2}$.
\end{conjecture}

\section{Bounding the entanglement for two-qubit states}\label{Sec:Bounds:Qubits}

To illustrate the idea of bounding entanglement directly
from measurement statistics, we now consider the specific case where the underlying quantum system can be described as a two-qubit state. It is worth noting that in this case, the entanglement of a quantum state
 $\rho$ is fully characterized in terms of its {\em entanglement of formation}, which is monotonically related to the {\em concurrence}, defined as~\cite{W.K.Wootters:PRL:2245}
\begin{equation}
    C=\max\left[0,2\sqrt{\lambda_1}-\sum_{i=1}^4\sqrt{\lambda_i}\right],
\end{equation}
where $\{\lambda_i\}$ are the decreasingly ordered
eigenvalues of
$\rho\,(\sigma_y\otimes\sigma_y)\rho\t(\sigma_y\otimes\sigma_y)$, $\sigma_y$ is the Pauli-$y$ matrix,
and the transposition $\t$ is to be carried out in any product
basis.

Now, let us imagine that in some experiments, the probability distributions $\AllProb$ are found.  From here, it is straightforward to compute the quantum value of $\St$ [cf.   Eq.~\eqref{Eq:Bellfunction:tiltedCH},] which we denote by $\StObs$. Clearly, if $\StObs$ corresponds to the maximal violation of $\It$ for some $\tau$, we can identify the concurrence of the underlying two-qubit state immediately with the help of Fig.~\ref{Fig:ConcurrencevsTau}. In general, there is of course no reason to expect that the measurement statistics will lead to such extremal correlations, in which case the bounds established below will be useful.

\subsection{Lower bound}
To see how a lower bound on the concurrence of $\rho$ can be obtained from  $\StObs$ --- in particular $\S^{(\tau=1)}_{\mbox{\tiny Obs}}=\SCH_{\mbox{\tiny Obs}}$ --- it suffices to recall from Ref.~\cite{F.Verstraete:PRL:170401} that for any two-qubit state $\rho$
\begin{equation}\label{Eq:UB:SCH}
	 \SqCH(\rho)\le\half\left(\sqrt{1+C^2}-1\right).
\end{equation}
It then follows from simple calculation that~\cite{fn:CHvsGeneralTau}
\begin{equation}\label{Eq:LB:C}
    C\ge\sqrt{\left(2\SqCH(\rho)+1\right)^2-1}
    \ge\sqrt{\left(2\SCHObs(\rho)+1\right)^2-1},
\end{equation}
where the second inequality follows from the fact that in general, $\SqCH(\rho)\ge\SCHObs(\rho)$, since the measurements operators $\{\GPOVMA,\GPOVMB\}$ may not be optimal. Also, it is worth noting that the first inequality in Eq.~\eqref{Eq:LB:C} can be saturated by pure states as well as some rank 2 mixed two-qubit states~\cite{F.Verstraete:PRL:170401}.

As a result of Eq.~\eqref{Eq:LB:C}, for any measurement statistics that violate inequality~\eqref{Ineq:tilted:CH}, that is, giving $\SCHObs(\rho)>0$, we know that its concurrence is bounded from below according to Eq.~\eqref{Eq:LB:C}.

\subsection{Upper bound}

\label{Sec:UB:BIV}

\begin{figure}[h!]
    \includegraphics[width=8cm]{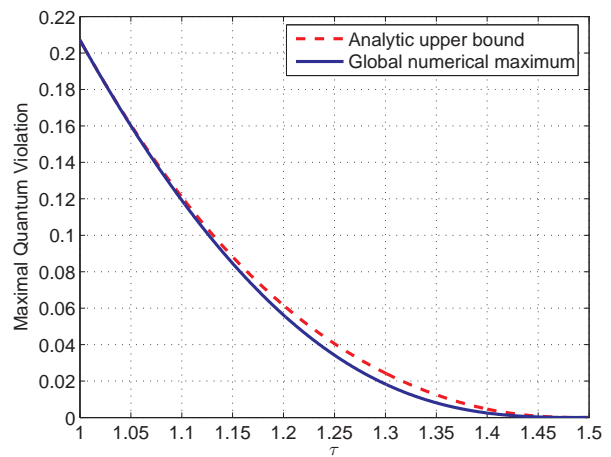}
    \caption{\label{Fig:MaximalBIV}
    (Color online) Maximal quantum violation of $I^{(\tau)}$
    as a function of $\tau$ (blue solid line). Also shown is the
    analytic upper bound on $\Sqt(\ket{\psi^*_\tau})$ computed from
    Eq.~\eqref{Eq:Sqm:AnalyticUB:PureState} (red dashed line), which
    provides an upper bound on $\Sqt$. }
\end{figure}

To obtain a semi-device-independent upper bound on the
concurrence of $\rho$, we make use of  $\Sqt(\rho)$ for a general two-qubit pure state $\rho$, cf.   Eq.~\eqref{Eq:MaxSqTau:FixedRho}, which can be determined easily using the tools from Ref.~\cite{Y.C.Liang:PRA:042103} and Ref.~\cite{QMP.Hierarchy}.
Specifically, if we denote a general two-qubit pure state (in the Schmidt basis) by
\begin{equation}\label{Eq:TwoQubit:Schmidt}
   \GPSqb=\cos\gamma\ket{0}_\A\ket{0}_\B +
    \sin\gamma\ket{1}_\A\ket{1}_\B,
\end{equation}
where $0\le\gamma\le\tfrac{\pi}{4}$, it can be shown that for any given $\gamma$, the maximal violation of $\GPSqb$ of $\It$ is upper bounded as follows (see Fig.~\ref{Fig:MaximalBIV})~\cite{fn:Analytic}:
\begin{equation}\label{Eq:Sqm:AnalyticUB:PureState}
    \Sqt(\ket{\psi^\gamma_2})\le\max
    \left\{0,2(1-\tau)\sin^2\gamma
    +\frac{\sqrt{1+\sin^22\gamma}-1}{2}\right\}.
\end{equation}
A direct consequence of this is that for any given $\gamma$, there exists a critical value of $\tau$, denoted by $\tau_c(\gamma)$ beyond which no violation of $\It$ by $\GPSqb$ is possible. Equivalently, for any given value of $\tau\ge \tau_c(\tfrac{\pi}{4})= \tfrac{1}{\sqrt{2}}+\frac{1}{2}$, there exists a critical state $\PsiCr= \GPSqb$  such that any state that is more entangled than $\PsiCr $ (i.e, one with a higher value of $\gamma$) cannot violate $\It$.  See Fig.~\ref{Fig:ConcurrencevsTau}.

From the above observation, we see that for any given $\AllProb$ that violates $\It$ for $\tau\ge \tau_c(\tfrac{\pi}{4})$, we can immediately place an upper bound on $\gamma$, and hence the corresponding concurrence
\begin{equation}\label{Eq:C:PureState}
	C_\gamma=\sin2\gamma,
\end{equation}
by first scanning through $\It$ for $\tau\ge \tau_c(\tfrac{\pi}{4})$ and determining the largest value of $\tau$ for which $\It$ is violated, that is,
\begin{equation}
	\tau_{\mbox{\tiny Obs}}=\sup \left\{\tau\in\left[ \frac{1}{\sqrt{2}}+\frac{1}{2},\frac{3}{2}\right] \,| \,\StObs>0\right\}.
\end{equation}
The concurrence of the two-qubit state that gives rise to $\AllProb$ is then upper bounded by $C\left(\ket{\psi^{\tau_{\mbox{\tiny Obs}}}_{\mbox{\tiny $c$}}}\right)$.

While the explicit value of $C$ corresponding to $\PsiCr$ for a given $\tau$ --- which we denote by $C_\text{cr}(\tau)$ --- can be easily determined using the tools from Ref.~\cite{Y.C.Liang:PRA:042103}, an analytic expression for $C_\text{cr}(\tau)$ has remained elusive. Nonetheless, from some simple calculation using Eq.~\eqref{Eq:Sqm:AnalyticUB:PureState}, one finds that $C_\text{cr}(\tau)$ must satisfy the following upper bound:
\begin{equation}\label{Eq:C:AnalyticUB:PureState}
    C_\text{cr}(\tau)\le C_\tau=
    \frac{2\sqrt{2(\tau-1)(2\tau-1)(3-2\tau)}}{5-8\tau+4\tau^2}.
\end{equation}

An important point to stress now is that although the upper bound obtained using the above procedures was derived from two-qubit pure state, the same bound is also applicable to any two-qubit mixed state $\rho$. This follows from the fact that any $\rho$ having concurrence $C$ can always be written as a convex decomposition of pure states, each having concurrence $C$~\cite{W.K.Wootters:PRL:2245}  and that $\Sqt(\rho)$ is linear in the convex decomposition of $\rho$.

Note also that stronger bounds can be obtained under the additional assumption that the measurements are projective (see Appendix~\ref{App:Projective}).

\subsection{An example}\label{Sec:Bounds:Example}

To see how these bounds work in practice, let us consider the following measurement statistics that were gathered~\cite{JD+Enrico} using a commercially available setup for producing a maximally entangled two-qubit state~\cite{fn:Uncertainty}.

\begin{gather}
	p(0,0|0,0)=0.3811,\quad p(0,0|1,0)=0.3789,\nonumber\\
	p(0,0|1,0)=0.3593,\quad p(0,0|1,1)=0.0671, \nonumber\\
	\MProbA{0}{0}=0.4025,\quad \MProbA{1}{0}=0.4806, \nonumber\\
	\MProbB{0}{0}=0.4671,\quad \MProbB{1}{0}=0.5058,
	\label{Eq:RealData}
\end{gather}
where $a,b=0$.

A simple calculation using Eq.~\eqref{Eq:LB:C} shows that the underlying two-qubit state that generates these probability distributions must have its concurrence lower bounded by 0.9297. Incidentally, these probability distributions also violate $\It$ up until $\tau_{\mbox{\tiny Obs}}\approx1.2102$. Using Eq.~\eqref{Eq:C:AnalyticUB:PureState}, this gives an upper bound on concurrence of 0.9999~\cite{fn:Diff}.  If we further make use of that fact the measurements that give rise to Eq.~\eqref{Eq:RealData} are projective [cf.   Appendix~\ref{App:Projective}], then this upper bound can be improved further via Eq.~\eqref{Eq:UB:Marginal} to 0.9806.

\section{Beyond qubits}\label{Sec:Qudits}

Let us now relax the assumption that the given quantum system (or more precisely, its degrees of freedom where the measurements were performed) are qubits and see what we can learn from the measurement statistics. In general, as one might expect, it is very difficult to derive any useful bounds on the entanglement of the underlying state if we are only given $\AllProb$. There are, nonetheless, exceptional cases where concrete statements can be made about the underlying state even in the scenario where we are only given a small set of independent numbers, such as  $\AllProb$.

The first example is when the measurement statistics give $\SqCH=\tfrac{1}{\sqrt{2}}-\tfrac{1}{2}$. In this case, the correlations between the measurement outcomes gives rise to the so-called Tsirelon bound~\cite{B.S.Tsirelson:LMP:1980}--- the maximal possible quantum violation of the CH (or CHSH) inequality. It was known since the early 1990s that in this case the underlying state must be even-dimensional, and is either a coherent superposition or an incoherent  mixture of two-qubit maximally entangled states that reside in orthogonal two-qubit subspaces~\cite{S.L.Braunstein:PRL:1992,S.Popescu:PLA:411}.

Likewise, if we denote by $\ket{\psi_\tau^*}$ the unique two-qubit pure state (in the form of $\GPSqb$) that maximally violates $\It$ (see Fig.~\ref{Fig:ConcurrencevsTau}), then from the work of Masanes~\cite{Ll.Masanes:PRL:050503} (see also Ref.~\cite{S.Pironio:NJP:diqkd}), one can show that the underlying quantum state must also be even-dimensional, and can either be a  coherent superposition or an incoherent  mixture of $\ket{\psi_\tau^*}$ that reside in orthogonal two-qubit subspaces.

Clearly, these examples indicate that when the dimension of the Hilbert space is not known, then even in these extremal cases, it is possible to have a wide spectrum of states, featuring different amounts of entanglement, that give rise to the same correlations. Hence, as long as we restrict ourself to the framework where only $\AllProb$ are given, it seems extremely challenging to construct any useful bounds on the underlying entanglement in a fully device-independent
manner.

\section{Conclusion}\label{Sec:Conclusion}

In this paper, we have investigated the possibility of bounding the entanglement of some underlying quantum state directly from measurement statistics, specifically from a given set of probability distributions $\AllProb$. Our main tool is a family of Bell-type inequalities whose maximal violation cannot be obtained from maximally entangled states of any dimension. Using these inequalities, we have illustrated how useful lower and upper bounds on the concurrence of some underlying two-qubit state may be constructed in a semi-device-independent
manner. In the case where the measurements are known to be projective, we also demonstrated how another upper bound can be obtained by analyzing the marginal probabilities given.

An interesting issue is to compare our approach with the one of Ref. \cite{DISE}, in which the authors investigate device-independent state estimation based on the CHSH inequality violation. The main differences between these two approaches are as follows. On the one hand, the approach of Ref. \cite{DISE} does not require any assumption about the Hilbert space dimension of the systems, in contrast to our approach. On the other hand, the observed CHSH violation $\SCHObs$ considered in Ref.~\cite{DISE} is actually $\SCHObs\left(\Lambda\left(\rho\right)\right)$, where $\Lambda$ refers to some local (nonunitary), completely positive map acting on the original state $\rho$, whereas in the present paper, we use directly $\SCHObs(\rho)$ to establish the bounds. Moreover, in the approach of Ref.~\cite{DISE}, the measurements that give rise to $\SCHObs\left(\Lambda\left(\rho\right)\right)$ are assumed to be optimal for $\Lambda\left(\rho\right)$, whereas we make no such assumption.

In general, it would be interesting to derive useful (semi-)device-independent bounds on entanglement for higher dimensional systems.
Admittedly, the Bell inequalities that we have considered here do not seem to provide useful bounds, as one may already anticipate from earlier work on dimension witnesses~\cite{DW}. However, given that there are other Bell inequalities~\cite{CGLMP, D.Collins:JPA:2004,Liang:PRA:2009,JD:JPA:2010} that are more suitable for higher dimensional quantum systems, it will be interesting to investigate if analogous bounds on, say, the generalized concurrence can be derived using these other inequalities.

As mentioned in Sec.~\ref{Sec:Prelim}, the Bell inequalities that we have considered here are closely related to those being used in the analysis of detection loopholes~\cite{Branciard:1010.1178,P.H.Eberhard:PRA:747,T.Vertesi:PRL:060401}. Another possible avenue for future research is to exploit results established in these earlier studies~\cite{Branciard:1010.1178,DetectionLoopholes} to establish or improve bounds on entanglement beyond what we have derived in this paper.

Finally, let us remark that for some quantum resources such as the $n$-partite Greenberger-Horne-Zeilinger state, it was shown that there is a high probability of finding Bell-inequality violations by performing local measurements in randomly chosen bases \cite{Liang:RFFBIV}. If we can also establish non-trivial bounds on multipartite entanglement  directly from these measurement statistics, then together, these approaches may offer a much more resource-economical paradigm for characterizing quantum states produced in the laboratory. From a quantum
information processing point of view, this is the ultimate goal that we hope to achieve with the approach taken in this paper. Note, however, that in the multipartite scenario, the possibility of bounding entanglement directly from measurement statistics may also become highly non-trivial, as there is no natural choice of entanglement measure, and the whole problem also becomes considerably more complicated.

\emph{Note added. While finishing this manuscript, we became aware of a related result by Vidick and Wehner \cite{Vidick:1011.5206}, who showed that the Bell inequality $I_{3322}$ of Ref. \cite{D.Collins:JPA:2004} shares a similar feature with the ones presented here, namely that its maximal quantum violation cannot be reached by maximally entangled states in any dimension. It is interesting to note that while our inequalities use only two measurements per party (compared to three measurements for $I_{3322}$), the inequality $I_{3322}$ is tight, in the sense of being a facet of the local polytope.}

\begin{acknowledgements}
The authors acknowledge helpful comments from an anonymous referee as well as useful discussions with Antonio Ac\'in, Jonathan Allcock, Joonwoo Bae, Jean-Daniel Bancal, Cyril Branciard,  Nicolas Gisin, Charles Lim, Enrico Pomarico, and Valerio Scarani. This work is supported by the Australian Research Council, the Swiss NCCR ``Quantum Photonics", the European ERC-AG QORE, the J\'anos Bolyai Grant of the Hungarian Academy of Sciences and the UK EPSRC.
\end{acknowledgements}

\appendix

\section{Proof of Theorem 1}\label{App:Proofs}

\subsection{Proof that there are $\AllProb$ that cannot be attained by any finite-dimensional maximally
entangled pure states}\label{App:MES:NMaxBIV}

Here, we provide a proof of Theorem~\ref{Conjecture:MESCannotProduceQ} which shows that there are bipartite probability distributions $\AllProb$ that cannot be written in the form of \begin{equation}
	\tr\left(\proj{\Psi^+_d}\GPOVMA\otimes\GPOVMB\right)
\end{equation}
where $\ket{\Psi^+_d}$ is the $d$-dimensional maximally entangled state. Note that similar results for more complicated experimental scenario were also obtained in Refs.~\cite{Junge:1007.3043,Vidick:1011.5206}.

\begin{proof}
The proof is by contradiction. Let us start by assuming the converse, namely, that the maximal
violation of $\It$ for $\tau\ge\tfrac{1}{\sqrt{2}}+\tfrac{1}{2}$ can also be
achieved using a $d$-dimensional maximally entangled state
\begin{equation}
    \ket{\Psi^+_d}=\frac{1}{\sqrt{d}}\sum_{j=1}^d \ket{j}_\A\ket{j}_\B,
\end{equation}
in conjunction with some choice of (non-trivial) projectors
$A_{\sA}^{\oA}$, $B_{\sB}^{\oB}$. We deduce the necessary
conditions that follow from this assumption and show that when these
necessary conditions are in place, $\ket{\Psi^+_d}$ cannot even
violate $\It$.

The assumption that $\rho=\proj{\Psi^+_d}$ can violate $\It$ maximally means that there must exist some choice of $A_{\sA}^{\oA}$ and $B_{\sB}^{\oB}$ such that [cf.   Eq.~\eqref{Eq:Bellfunction:tiltedCH}]
\begin{subequations}
\begin{align}
    \Sqt&= \Sqt\left(\rho,\left\{A_{\sA}^{\oA}, B_{\sB}^{\oB}\right\}\right),\label{Eq:Sq.tau1}\\
    &=\sum_{\sA,\sB=0,1}\sum_{\oA,\oB=0,1} \beta^{\oA\oB}_{\sA\sB}\tr\left(\rho\, A_{\sA}^{\oA}\otimes B_{\sB}^{\oB}\right).\label{Eq:Sq.tau2}
\end{align}
\end{subequations}
Without loss of generality, we now make use of the main lemma from Ref.~\cite{Ll.Masanes:PRL:050503} and assume that the state $\rho$ and all the POVM elements are already written in the basis whereby all the four $A_{\sA}^{\oA}$ are simultaneously block diagonal, likewise for $B_{\sB}^{\oB}$, that is,
\begin{align}
    A_{\sA}^{\oA}&=\sum_i P_{\A,i} A_{\sA}^{\oA} P_{\A,i}=\bigoplus_i A_{\sA,i}^{\oA},\nonumber\\
    B_{\sB}^{\oB}&=\sum_i P_{\B,i} B_{\sB}^{\oB} P_{\B,i}=\bigoplus_i B_{\sB,i}^{\oB},
\end{align}
where $P_{\A,i}$, $P_{\B,i}$ are (local) block-diagonal projectors consisting of only $2\times2$ and/or
$1\times1$ blocks.

Let us now denote by $\sigma_{ij}$ the (unnormalized) non-vanishing $4\times4$,
$2\times2$, or $1\times1$ blocks of $P_{\A,i}\otimes P_{\B,j}\,\rho\,P_{\A,i}\otimes P_{\B,j}$ and $p_{ij}$ the corresponding probability of projecting into this block, that is, $p_{ij}=\tr P_{\A,i}\otimes P_{\B,j}\,\rho\,P_{\A,i}\otimes P_{\B,j}$. The normalized density matrix corresponding to $\sigma_{ij}$ can then be written as $\rho_{ij}=\sigma_{ij}/p_{ij}$. In these notations, Eq.~\eqref{Eq:Sq.tau2} can be rewritten as
\begin{align*}
    \Sqt&=\sum_{i,j}\,^{'}p_{ij} \sum_{\sA,\sB=0,1}\sum_{\oA,\oB=0,1} \Coeff
    \tr\left(\rho_{ij}\, A_{\sA,i}^{\oA}\otimes B_{\sB,j}^{\oB}\right),\\
    &=\sum_{i,j}\,^{'}p_{ij}\, \Sqt\left(\rho_{ij},\left\{A_{\sA,i}^{\oA}, B_{\sB,j}^{\oB}\right\}\right),
\end{align*}
where the restricted sum $\sum'$ only runs over $i$ and $j$ where $p_{ij}\neq0$, or equivalently where
$\rho_{ij}$ is well defined. This last expression indicates
that $\Sqt$ can also be seen as a
convex combination of the quantum values of $\rho_{ij}$ under
the POVM elements $\{A_{\sA,i}^{\oA}\}$, $\{B_{\sB,j}^{\oB}\}$.
Since $\Sqt$ is the maximal quantum violation of $\It$, so we must also have
\begin{equation}\label{Eq:ConvexSum=Full}
    \Sqt=\Sqt\left(\rho_{ij},\left\{A_{\sA,i}^{\oA}, B_{\sB,j}^{\oB}\right\}\right)
\end{equation}
for all $i,j$ where $\rho_{ij}$ is non-vanishing.

Clearly, in order for Eq.~\eqref{Eq:ConvexSum=Full} to hold true, for any values of $i,j$ where $p_{ij}\neq0$, the corresponding $\rho_{ij}$ cannot be a block of size $2\times2$ or $1\times1$, since such blocks correspond to classical states that cannot violate $\It$, and hence do not satisfy Eq.~\eqref{Eq:ConvexSum=Full}. This implies that (i) $\ket{\Psi^+_d}$ cannot have odd Schmidt rank and (ii) the projectors $P_{\A,i}$, $P_{\B,j}$ cannot contain any nonvanishing $1\times1$ block, or otherwise there is always a nonzero probability of projecting onto a $\rho_{ij}$ with size $2\times2$ or smaller. Combining these observations with the main lemma of Ref.~\cite{Ll.Masanes:PRL:050503} gives $\tr(A_{\sA,i}^{\oA})=\tr(B_{\sB,j}^{\oB})=1$ for all $x,y,a,b,i,j$, and hence
\begin{equation}
\tr(A_{\sA}^{\oA})=\tr(B_{\sB}^{\oB})=\frac{d}{2}.
\end{equation}
In other words, the marginal probabilities become
\begin{equation*}
    \GMProbA=\tr(\proj{\Psi^+_d}A_{\sA}^{\oA}\otimes\unit_\B)
    =\frac{\tr(A_{\sA}^{\oA})}{d}=\frac{1}{2}\quad\forall\,a,x,
\end{equation*}
and likewise for $\GMProbB$.
Putting these back into Eq.~\eqref{Eq:Bellfunction:tiltedCH:ViaCH}, we see that for an even-dimensional
maximally entangled state and where Eq.~\eqref{Eq:ConvexSum=Full} is enforced, its quantum value becomes
\begin{equation}\label{Eq:SqmMES:constrained}
    \Sqt\left(\rho,\{\GPOVMA,\GPOVMB\}\right)=\SqCH\left(\rho,\{\GPOVMA,\GPOVMB\}\right) -(\tau-1).
\end{equation}
The first part of this expression is simply the ordinary CH
polynomial, which admits the maximum quantum value
$\tfrac{1}{\sqrt{2}}-\tfrac{1}{2}$ for any even-dimensional
maximally entangled state~\cite{N.Gisin:PLA:15,Y.C.Liang:PRA:052116}. Is is then easy to see that the left-hand side of Eq.~\eqref{Eq:SqmMES:constrained} is always upper bounded by 0 for
\begin{equation}
    \tau\ge\frac{1}{2}+\frac{1}{\sqrt{2}}\approx1.2071,
\end{equation}
which contradicts our initial assumption that $\ket{\Psi^+_d}$ can violate $\It$ maximally for this interval of $\tau$.
\end{proof}

\section{Upper bounding entanglement assuming that the measurements are projective}
\label{App:Projective}

Is it still possible to upper bound the entanglement of the underlying state when the probability distributions $\AllProb$ do not violate $\It$ for $\tau>\tau_c(\tfrac{\pi}{4})$? In this regard, it turns out that a simple upper bound on $C$ can always be established if we are also equipped with the knowledge that the measurements that give rise to $\AllProb$ are projective measurements, that is, are described by
\begin{equation}\label{Eq:TwoOutomceProjectors}
    A^{\oA}_{\sA}=\half\left[\unit_2 +(-1)^{\oA}\hat{a}_{\sA}
    \cdot\vec{\sigma}\right],
    \quad
    B^{\oB}_{\sB}=\half\left[\unit_2 +(-1)^{\oB}\hat{b}_{\sB}
    \cdot\vec{\sigma}\right],
\end{equation}
where $\vec{\sigma}$ is the vector of Pauli matrices, $\hat{a}_x, \hat{b}_y\in\mathbb{R}^3$ are unit vectors, and $\unit_2$ is the $2\times2$ identity matrix.

The key idea here is to realize that as the concurrence $C$ increases from 0 to 1 (i.e., $\gamma$ increases from 0 to $\frac{\pi}{4}$), the interval of admissible marginal probabilities $\GMProbA$, $\GMProbB$ shrinks from $[0,1]$ to a single point $\frac{1}{2}$. This can be seen, for example, by noting that for $\GPSqb$ and with Eq.~\eqref{Eq:TwoOutomceProjectors}, we have
\begin{align}\label{Eq:MarginalProb:Schmidt}
	\GMProbA&=\frac{1}{2}(1+\cos\theta_x\cos2\gamma), \nonumber\\
	\GMProbB&=\frac{1}{2}(1+\cos\theta_y\cos2\gamma),
\end{align}
where $\cos\theta_x$ ($\cos\theta_y$) is the $z$-component of the vector $\hat{a}_x$ ($\hat{b}_y$). Clearly, these expressions are bounded as
\begin{align}\label{Eq:MarginalProb:Bound}
    \frac{1}{2}(1-\cos2\gamma)\le \GMProbA,\, \GMProbB \le \frac{1}{2}(1+\cos2\gamma).
\end{align}
Together with Eq.~\eqref{Eq:C:PureState}, it is then easy to see that
\begin{equation}\label{Eq:UB:Marginal}
	C \le \min_{a,b,x,y} \left\{ \sqrt{1-\left[1-2\GMProbAObs\right]^2},\sqrt{1-\left[1-2\GMProbBObs\right]^2}\right\},
\end{equation}
where $\GMProbAObs$ and $\GMProbBObs$ are, respectively, Alice's and Bob's marginal probabilities estimated from the experiment.

It is important to note that although Eq.~\eqref{Eq:MarginalProb:Bound} was derived for  $\GPSqb$, it also holds for an arbitrary pure two-qubit state having the same concurrence. Consequently, as with the upper bound derived in the last section, Eq.~\eqref{Eq:UB:Marginal} is also applicable to arbitrary two-qubit mixed states.



\begin{thebibliography}{99}
\bibitem{fn:Filters}  Naively, one might expect that by first grouping the local subspaces of some higher-dimensional maximally entangled state, it is possible to deterministically produce a two-qubit non-maximally entangled state which can then be used to give maximal violation of $\It$. However, it turns out that such operation, which consists in diluting entanglement, cannot be achieved without classical communication \cite{Aram.Harrow:CC}.
\bibitem{fn:DetectionLoophole} Note that if this conjecture is valid, then no finite-dimensional
maximally entangled states with a symmetric detection efficiency of
$\left(\tfrac{1}{\sqrt{2}}+\tfrac{1}{2}\right)^{-1}\approx82.8\%$ or less can generate $\AllProb$ that
could be used to close the detection loophole.
\bibitem{fn:CHvsGeneralTau} Note that inequality~\eqref{Eq:LB:C} is also valid with $\SqCH$ and $\SCHObs$ replaced, respectively, by $\Sqt$ and $\St_{\mbox{\tiny Obs}}$. However, it should be clear from Eq.~\eqref{Eq:Bellfunction:tiltedCH:ViaCH} that the best lower bound on $C$ that can be obtained from the measurement statistics is due to the one where $\tau=1$, in which case $\Sqt$ becomes $\SqCH$ and $\St_{\mbox{\tiny Obs}}$ becomes $\SCHObs$.
\bibitem{fn:Uncertainty} For simplicity, we will omit the (approximately) 1\% uncertainty in the presentation of the measurement statistics.
\bibitem{fn:Diff} It turns out that in this case, the difference between $C_\tau$ and $C_{\mbox{\tiny Cr}}$ is less than $10^{-5}$.
\end{thebibliography}
\end{document}